\documentclass[conference]{IEEEtran}
\IEEEoverridecommandlockouts
\usepackage{cite}
\usepackage{amsmath,amssymb,amsfonts}
\usepackage{algorithmic}
\usepackage{graphicx}
\usepackage{textcomp}
\usepackage{xcolor}

\usepackage{comment}
\usepackage{tikz}
\usetikzlibrary{positioning, arrows.meta}
\usepackage{listings}

\def\BibTeX{{\rm B\kern-.05em{\sc i\kern-.025em b}\kern-.08em
    T\kern-.1667em\lower.7ex\hbox{E}\kern-.125emX}}

\usetikzlibrary{positioning, arrows.meta}

\definecolor{codegreen}{rgb}{0,0.6,0}
\definecolor{codegray}{rgb}{0.5,0.5,0.5}
\definecolor{codepurple}{rgb}{0.58,0,0.82}
\definecolor{backcolour}{rgb}{0.95,0.95,0.92}

\lstdefinestyle{mystyle}{
    backgroundcolor=\color{backcolour},   
    commentstyle=\color{codegreen},
    keywordstyle=\color{magenta},
    numberstyle=\tiny\color{codegray},
    stringstyle=\color{codepurple},
    basicstyle=\ttfamily\footnotesize,
    breakatwhitespace=false,         
    breaklines=true,                 
    captionpos=t, 
    keepspaces=true,                 
    numbers=left,                    
    numbersep=5pt,                   
    showspaces=false,                
    showstringspaces=false,
    showtabs=false,                  
    tabsize=2,
    xleftmargin=17pt,               
    framexleftmargin=17pt,          
}

\begin{document}

\title{Natural Language to Verilog: Design of a Recurrent Spiking Neural Network using Large Language Models and ChatGPT
\\
}

\author{\IEEEauthorblockN{Paola Vitolo\IEEEauthorrefmark{1}\IEEEauthorrefmark{2},
George Psaltakis\IEEEauthorrefmark{1}\IEEEauthorrefmark{3},
Michael Tomlinson\IEEEauthorrefmark{1},
Gian Domenico Licciardo\IEEEauthorrefmark{2},
and Andreas G. Andreou\IEEEauthorrefmark{1}
}
\IEEEauthorblockA{\IEEEauthorrefmark{1}Department of Electrical \& Computer Engineering,\\Johns Hopkins University, Baltimore, MD, USA\\
Email: \{pvitolo1, gpsalta1, mtomlin5, andreou\}@jhu.edu}
\IEEEauthorblockA{\IEEEauthorrefmark{2}Department of Industrial Engineering\\
University of Salerno, Fisciano, IT}
\IEEEauthorblockA{\IEEEauthorrefmark{3}Department of Electrical \& Computer Engineering,\\Hellenic Mediterranean University, Heraklion, GR}
}
\maketitle

\begin{abstract}
This paper investigates the use of Large Language Models (LLMs) and natural language prompts to generate  hardware description code, namely Verilog.
Building on our prior work, we employ OpenAI's ChatGPT4 and natural language prompts to synthesize an RTL Verilog module of a programmable recurrent spiking neural network, while also generating test benches to assess the system's correctness.
The resultant design was validated in three simple machine learning tasks, the exclusive OR, the IRIS flower classification and the MNIST hand-written digit classification. Furthermore, the design was validated on a Field-Programmable Gate Array (FPGA) and subsequently synthesized in the SkyWater 130 nm technology by using an open-source electronic design automation flow. The design was submitted to Efabless Tiny Tapeout 6.




\end{abstract}

\begin{IEEEkeywords}
generative AI, LLM, neuromorphic architecture, open-source hardware design, recurrent spiking neural network.
\end{IEEEkeywords}

\section{Introduction}
In recent years, the demand for custom integrated circuits has increased exponentially, driving the Application-Specific Integrated Circuits (ASICs) market with estimates reaching the USD 33.3 billion by 2033, with an annual growth rate of 6.4\% \cite{2024_asic_market_us_full_report}.
Industries, including IoT, automotive, consumer electronics, and healthcare, increasingly require tailored solutions with custom ASICs to improve their products' in most cases adding AI/ML capabilities\cite{2024_asic_market_us_full_report, 2018_stethoVest,2023_asic_in_MEMS,2018_micro_doppler_sonar,2023_asic_for_KWS,hybrid}.
Lengthy design and verification times, combined with the increasingly stringent demands for ``time to market'' pose significant challenges in the hardware design process \cite{2018_chip_development}.

To address these challenges, generative Artificial Intelligence (GenAI) and specifically Large Language Models (LLMs), could augment state-of-the-art Electronic Design Automation (EDA) tools, and offer viable solutions \cite{2021_eda_ML_survey}.
Specifically, in this paper we demonstrate that it is possible to use natural language prompts to generate hardware description code, namely Verilog. 
Since the advent of LLMs in the 4Q of 2023, there has been efforts to use LLMs in the hardware design to streamline and accelerate the process \cite{2023_VerilogEval_LLM, 2024_LLM_for_Verilog,2024_Hardware_Phi_LLM_for_HW_design_and_verification,2023_Benchmarking_LLM_for_RTL}. 
J. T. Meech in \cite{2023_RNG_LLM} used Microsoft Bing-chat in order to generate a permuted congruential random number generator design with a wishbone interface.
Tailoring for hardware debugging, Fu et al. used them for the identification and resolutions of bugs in hardware designs \cite{2023_LLM4SecHW_LLM_for_HW_design_and_verification}.
The most popular choice for LLM based hardware design has been OpenAI's GPT.
The authors in \cite{2023_chipChat, 2024_soft_hard_codesign_LLM } employed GPT to generate hardware description code in Verilog or VHDL, resulting in a reduction of syntax errors.
Similarly, our previous work tackled a programmable 3 by 3 digital spiking neuron array created from conversational language using GPT-4. The above implementation was submitted for fabrication in Skywater 130 nm technology and was verified through simulation \cite{2024_Mike_chatgptSNN}. 

While all of above showcase a promising shift in EDA technology there is still a need for improvements and further investigations to support more complex designs. 

In this paper we build on our prior work \cite{2024_Mike_chatgptSNN}, and explore GPT-4’s potential to design a more complex hardware design architecture, namely a 3 layer, 3 neuron per layer Recurrent Spiking Neuron Network (RSNN). Furthermore we demonstrate how to use natural language prompts and LLMs to support test-bed development and hence the verification process.

Three simple case studies of AI/ML were employed to validate the design, the exclusive OR, the IRIS flower classification, and the MNIST handwritten digit classification. The LLM generated Verilog code, was validated on a Field-Programmable Gate Array (FPGA) and implemented  Skywater 130 nm technology, utilizing an open source EDA tool.
The design has synthesized into an ASIC (that includes an SPI interface) using an open-source electronic design automation flow and was submitted to Efabless Tiny Tapeout 6.

\section{Methods}

{\bf Why SNNs?} Spiking Neural Networks (SNNs) are known as the third generation of artificial neural networks. They mimic the spiking behavior of biological neural networks and when mapped on silicon yield smaller architectures than traditional neural networks \cite{edgesnn,cassidy_design_2013}. More important, by leveraging recurrent connections, RSNNs show inherent capabilities in handling temporal dynamic series classification problems \cite{Spatiotemporal,craley_action_2017} and hence are well matched to real-time AI/ML tasks that necessitate low latency.

{\bf Design methodology}
We employ OpenAI's ChatGPT-4 and we follow a modular approach with a bottom-up design methodology. The architecture of the complex RSNN system is decomposted into a {\it hierarchy} into smaller, reusable submodules, which leads to speeding up the development process, improving manageability, scalability, reusability and facilitating early error detection.

ChatGPT was requested to describe each submodule in the hardware description language Verilog and generate the corresponding documentation, crucial for future code maintenance and upgrades.
Once a component was ready, ChatGPT generated the relative test bench to verify each module independently.
The final step was the instatiation all the modules in a top module, thereby constructing the entire system.

All source code and conversation transcripts are available in the publicly accessible Github repository at Andreou-JHULabOrg/tinytapeout\_06\_chatgpt\_rsnn \cite{tinytapeout_06_chatgpt_rsnn}.

\subsection{Implementation Details}

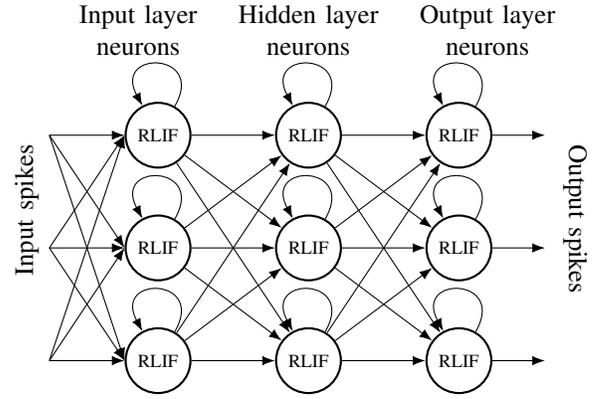
\begin{figure}[htbp]
\centering
\begin{tikzpicture}[
    node distance=3cm and 1.5cm, 
    neuron/.style={circle,draw,thick,minimum size=5mm,font=\scriptsize}, 
    arrow/.style={-Latex},
    label/.style={font=\footnotesize,rotate=90} , 
    inputlabel/.style={rotate=90},
    outputlabel/.style={rotate=-90}
]

\foreach \i in {1,...,3}
    \node[neuron] (Input-\i) at (0,-\i*1.5) {RLIF}; 

\foreach \i in {1,...,3}
    \node[neuron] (Hidden-\i) at (2,-\i*1.5) {RLIF}; 

\foreach \i in {1,...,3}
    \node[neuron] (Output-\i) at (4,-\i*1.5) {RLIF}; 

\foreach \i in {1,...,3}
    \foreach \j in {1,...,3}
        \draw[arrow] (Input-\i) -- (Hidden-\j);

\foreach \i in {1,...,3}
    \foreach \j in {1,...,3}
        \draw[arrow] (Hidden-\i) -- (Output-\j);

\foreach \i in {1,...,3}
{
    \draw[arrow] (Input-\i) to [out=60,in=118,looseness=5] (Input-\i);
    \draw[arrow] (Hidden-\i) to [out=60,in=118,looseness=5] (Hidden-\i);
    \draw[arrow] (Output-\i) to [out=60,in=118,looseness=5] (Output-\i);
}

\node[align=center, above=0.5cm of Hidden-1] (hidden label) {Hidden layer\\neurons}; 
\node[align=center, left=0.3cm of hidden label] {First layer\\neurons}; 
\node[align=center, right=0.3cm of hidden label] {Output layer\\neurons}; 

\foreach \i in {1,...,3}
{
    \draw[arrow] ([xshift=-1cm]Input-\i.west) -- (Input-\i.west);
    \draw[arrow] (Output-\i.east) -- ([xshift=0.7cm]Output-\i.east);
}

\draw[arrow] ([xshift=-1cm]Input-1.west) --(Input-2.west);
\draw[arrow] ([xshift=-1cm]Input-1.west) -- (Input-3.west);
\draw[arrow] ([xshift=-1cm]Input-2.west) --(Input-1.west);
\draw[arrow] ([xshift=-1cm]Input-2.west) -- (Input-3.west);
\draw[arrow] ([xshift=-1cm]Input-3.west) --(Input-2.west);
\draw[arrow] ([xshift=-1cm]Input-3.west) -- (Input-1.west);

\node[inputlabel, left=1.3cm of Input-1] {Input spikes};

\node[outputlabel, right=5.1cm of Input-1] {Output spikes};

\end{tikzpicture}
\caption{Schema of the desired Recurrent Spiking Neural Network, consisting of 3 fully connected layers, each layer having 3 recurrent spiking neurons.}
\label{fig:Fig1}
\end{figure}

The architecture of the synthesized RSNN is shown in Fig. \ref{fig:Fig1}.
The architecture consists of a fully connected neural network comprising three spike based inputs and three spike based outputs.
Hence there are 3 layers, each consisting of 3 recurrent spiking neurons. Each neuron of a layer is connected to all neurons of the previous layer, or to all its inputs in the case of the first layer, resulting in 3 weights per neuron. Consequently, each layer has a total of 3$\times$3 weights, amounting 3$\times$3$\times$3=27 weights across the entire network.

The equations describing the neuron model is that of a Leaky Integrate-and-Fire (LIF) recurrent spiking neuron modeled by \ref{eq:LIF_RSN}, where $I_{in}$ represents the input current, $U$ is the membrane potential, $U_{thr}$ denotes the membrane threshold, $S_{out}$ is the output spike, $R$ represents the reset mechanism, $\beta$ is the membrane potential decay rate, and $V$ is the feedback scale factor.
\begin{equation}
\label{eq:LIF_RSN}
\begin{aligned}
& U(t) > U_{thr}  \rightarrow S(t+1)=1\\
& U(t+1)=\beta U(t)+I_{in}(t+1)+VS_{out}(t)-RU_{thr}
\end{aligned}
\end{equation}
When the membrane potential exceeds $U_{thr}$, an output spike $S_{out}$ is generated and $R$ becomes 1, resetting the potential $U$ by subtracting the threshold value.
The recurrent mechanism is characterized by one-to-one connections, where each neuron is connected back to itself, scaling its output spike by V.
Additionally, the model includes a refractory period during which the membrane potential $U$ is not affected by neuron inputs after spiking. This feature ensures that the neuron does not immediately fire again, providing a more realistic emulation of neuronal behaviour.


Following a bottom-up design approach, the LIF spiking neuron was synthesized first. The original design produced by ChatGPT was iteratively refined  progressively detailing the specifications and systematically addressing the emerging errors.
The final LIF module provided by ChatGPT is presented in Table \ref{table:LIF_module}, alongside the initial output. Subsequently, ChatGPT produced the documentation for this module.


\begin{table*}[!htbp]
\centering
\caption{Leaky Integrate-and-Fire Spiking Neuron Verilog Module generated by ChatGPT}
\label{table:LIF_module}
\begin{minipage}[t]{.49\textwidth}
\begin{lstlisting}[language=Verilog, caption={Initial Module}]
module LeakyIntegrateFireNeuron(
    input clk,
    input reset_n,
    input [7:0] input_current,
    input [7:0] threshold,
    input [7:0] leak,
    input [3:0] refractory_period,
    output reg fired
);
    // Membrane potential of the neuron
    reg [7:0] membrane_potential;
    // Membrane potential of the neuron
    reg [3:0] refractory_counter;

    always @(posedge clk or negedge reset_n) begin
        if (!reset_n) begin
            membrane_potential <= 8'b0;
            refractory_counter <= 4'b0;
            fired <= 1'b0;
        end else begin
            // If in refractory period,
            // decrement the refractory counter
            if (refractory_counter > 0) begin
                refractory_counter <= refractory_counter - 1;
            end else begin
                // Update membrane potential
                // with leaky integration
                membrane_potential <= membrane_potential + input_current - (membrane_potential * leak >> 8);

                // Check if the membrane
                // potential has reached
                // or exceeded the threshold
                if (membrane_potential >= threshold) begin
                    // If threshold is
                    // reached, neuron fires
                    fired <= 1'b1;
                    // Reduce membrane
                    // potential by the
                    // threshold value
                    membrane_potential <= membrane_potential - threshold;
                    // Set refractory counter
                    // to the specified
                    //refractory period
                    refractory_counter <=
                        refractory_period;
                end else begin
                    // If threshold is not
                    // reached, neuron does
                    // not fire
                    fired <= 1'b0;
                end
            end
        end
    end
endmodule
\end{lstlisting}
\end{minipage}
\hfill
\begin{minipage}[t]{.49\textwidth}
\begin{lstlisting}[language=Verilog, caption={Final Module}]
module LeakyIntegrateFireNeuron(
    input clk,
    input reset,
    input enable,
    input [7:0] input_current,
    input [7:0] threshold,
    input [7:0] decay,
    input [7:0] refractory_period,
    output reg spike_out
);

    reg [7:0] membrane_potential = 8'b0;
    reg [7:0] refractory_counter = 8'b0;
    wire signed [9:0] potential_update;
    
    // Calculate the potential update in signed arithmetic based on input current and decay    
    assign potential_update = $signed({membrane_potential[7], membrane_potential[7], membrane_potential}) +$signed({input_current[7], input_current[7], input_current}) +(membrane_potential[7] ? $signed({decay[7], decay[7], decay}) : -$signed({decay[7], decay[7], decay}));

    always @(posedge clk or posedge reset) begin
        spike_out <= 1'b0;    
        if (reset) begin
        membrane_potential <= 8'b0;
        refractory_counter <= 8'b0;
        end else if (enable) begin
        if (refractory_counter > 0) begin
        refractory_counter <= refractory_counter - 1;
        end else begin
        // Check for potential overflow
        if (potential_update[9] && potential_update < -128)
        // If overflow in the negative direction,
        // saturate to minimum value
        membrane_potential <= 8'b1000_0000;
        else if (potential_update > 127)
        // If overflow in the positive direction,
        // saturate to maximum value
        membrane_potential <= 8'b0111_1111;
        else
        // If no overflow, assign the
        // 8 LSBs of updated potential
        membrane_potential <= potential_update[7:0];
        // Check if membrane potential exceeds
        // the threshold
        if ($signed(membrane_potential) >= $signed(threshold)) begin
        // Neuron fires
        spike_out <= 1'b1;
        // Reduce membrane potential
        // by the threshold value
        membrane_potential <= $signed(membrane_potential) - $signed(threshold);
        // Set refractory counter to the
        // specified refractory period
        refractory_counter <= refractory_period;
        end
        end
        end
    end
endmodule
\end{lstlisting}
\end{minipage}

\end{table*}

\begin{table*}[!htbp]
\centering
\caption{Test Bench for the LIF Spiking Neuron Verilog Module generated by ChatGPT}
\label{table:LIF_module_tb}

\begin{minipage}[t]{.49\textwidth}
\begin{lstlisting}[language=Verilog, caption={Initial Module}]
`timescale 1ns / 1ps
module tb_LeakyIntegrateFireNeuron;
    reg clk;
    reg reset;
    reg [7:0] input_current;
    reg [7:0] threshold;
    reg [7:0] decay;
    reg [7:0] refractory_period;
    wire spike_out;
    LeakyIntegrateFireNeuron uut (
        .clk(clk),
        .reset(reset),
        .input_current(input_current),
        .threshold(threshold),
        .decay(decay),
        .refractory_period(refractory_period),
        .spike_out(spike_out)
    );
    initial begin
        clk = 0;
        forever #10 clk = !clk;
    end
    initial begin
        reset = 1;
        input_current = 0;
        threshold = 50; 
        decay = 5;   
        refractory_period = 5;
        #25 reset = 0;
        // Stimulus 1:
        // Small input current, no spike
        input_current = 10;
        #40;
        // Stimulus 2: Larger current
        // to trigger spike
        input_current = 60;
        #20;
        // Stimulus 3: Check refractory
        // behavior, should not spike
        // despite high current
        input_current = 60;
        #100;
        // Stimulus 4: After refractory
        // period, should be able
        // to spike again
        input_current = 60;
        #100;
        // Complete the test
        #50 $finish;
    end
    initial begin
        $monitor("Time = %t, Input Current = %d, Membrane Potential = %d, Spike = %b", $time, input_current, uut.membrane_potential, spike_out);
    end
endmodule
\end{lstlisting}
\end{minipage}
\hfill
\begin{minipage}[t]{.49\textwidth}
\begin{lstlisting}[language=Verilog, caption={Final Module}]
`timescale 1ns / 1ps
module tb_LeakyIntegrateFireNeuron;
    reg clk;
    reg reset;
    reg enable;
    reg [7:0] input_current;
    reg [7:0] threshold;
    reg [7:0] decay;
    reg [7:0] refractory_period;
    wire spike_out;
    LeakyIntegrateFireNeuron uut (.clk(clk),.reset(reset), .enable(enable),.input_current(input_current),.threshold(threshold), .decay(decay), .refractory_period(refractory_period),.spike_out(spike_out));
    initial begin
        clk = 0;
        forever #10 clk = !clk;
    end
    initial begin
        reset = 1; enable = 0;
        input_current = 0; threshold = 127;
        decay = 1; refractory_period = 10;
        #25 reset = 0; enable = 1;
        input_current = 10;  #40;
        input_current = 60;  #20;
        input_current = 60;  #100;
        enable = 0;
        input_current = 60;  #40;
        enable = 1;          #20;
        input_current = 30;  #100;
        enable = 0;          #40;
        enable = 1;
        input_current = 75;  #60;
        reset = 1;           #30;
        reset = 0;           #40;
        input_current = 15;  #100;
        enable = 0;          #40;
        enable = 1;          #60;
        input_current = 5;
        $display("Starting test with low input and high threshold at time %t", $time);
        fork
            begin
                wait (spike_out == 1);
                $display("Spike occurred at time %t with low input current", $time);
                $finish;
            end
            begin
                #10000;
                $display("No spike after 10000ns with low input current");
                $finish;
            end
        join
    end
    initial begin
        $monitor("Time = %t, Input Current = %d, Membrane Potential = %d, Spike = %b", $time, input_current, uut.membrane_potential, spike_out);
    end
endmodule

\end{lstlisting}
\end{minipage}

\end{table*}

\begin{figure*}[htbp]
\centerline{\includegraphics[width=6in ]{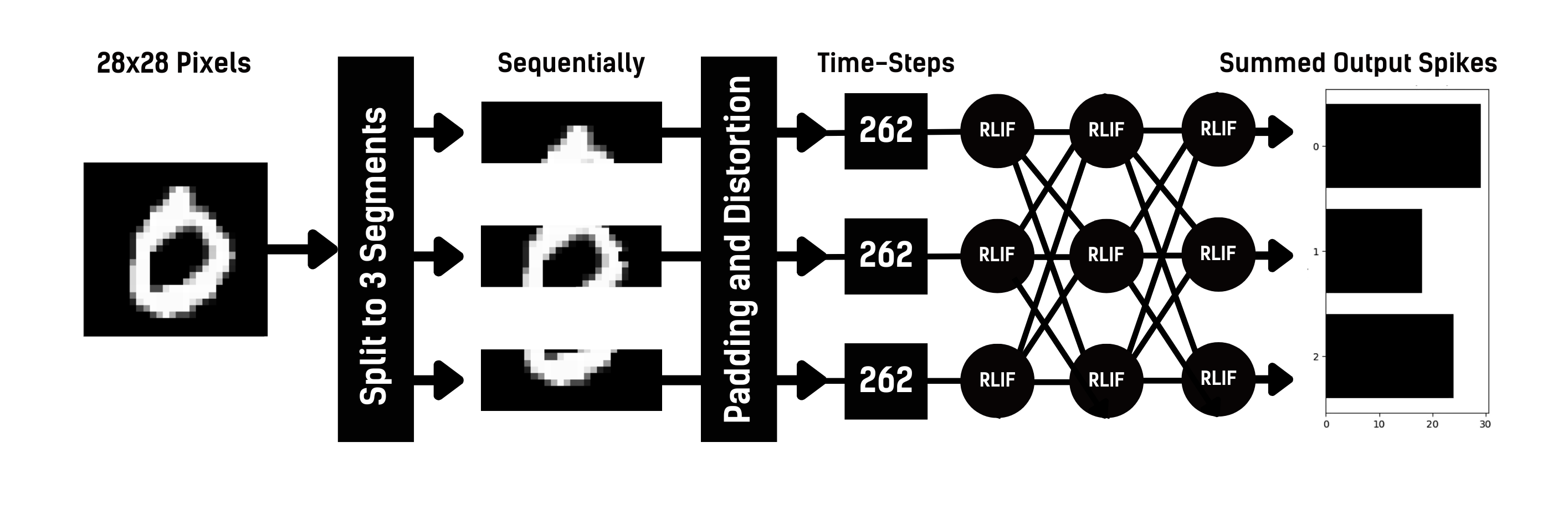}}
\caption{Block Diagram describing the pipeline used for our proposed Sequential MNIST model.}
\label{fig:neural1}
\end{figure*}

Using the same iterative  methodology, additional input for the feedback scale factor was included in the design. 
This was followed by an implementation of a layer of three neurons, managing the generation of input currents for each neuron by multiplying input spikes with weights.
Subsequently, a network with three layers, a First-In-Parallel-Out (FIPO) based memory for the neuron parameters, and a Finite State Machine (FSM)-based control unit to manage the memory operations were generated. The final design is the top module that integrates the network, memory, control unit for the memory, synchronizers for asynchronous inputs, and an array of state holdiing elements for the input spikes.

The design requirements included the system support both negative and positive weights, two's complement 8-bit fixed point encoding, management of both negative and positive overflow, and extension of arithmetic bit length to adequately support accumulation operations.


\subsection{Test Bench Generation with ChatGPT}
Once a module was ready, we opened a new chat conversation in a separate window prompted ChatGPT to generate a Verilog test bench for that module. In this task, the test-bench outputs were iteratively refined by troubleshooting and providing it more details for the test vectors. Each module was tested individually and, finally, all together with the top module.
Table \ref{table:LIF_module_tb} illustrates an example of the initial (left) and final (right) outputs of the LIF module test bench.

\section{ML/AI Validation Case Studies}
To assess the functional correctness of the Verilog modules developed by ChatGPT, we tested the generated design in three case studies: exclusive OR, IRIS flower classification, and the MNIST handwritten digit classification. SNNTorch\cite{jason} as well as the the documentation provided for the recurrent leaky integrate and fire neurons with one-to-one connections was used for modeling our design. The weights and the parameters of the neural network were quantized  using the libraries of snntorch\cite{jason} and brevitas\cite{brevitas}. Our architecture consisted of a three layer network with each of the layers having  a quantized linear layer with recurrent spiking neurons with one-to-one connections consisting of three neurons each. 

\subsection{Exclusive OR}
The exclusive OR is a simple classification problem to verify if the network can learn nonlinear boundaries. For this problem  only two of the three input neurons as well as only two of the three output neurons were used. Accuracies of up to  95\% were reached in this task.

\subsection{IRIS flower classification}
The IRIS dataset is a good match of a classification task for the synthesized RSSN architecture. The task consists of classifying the eata in the tree classes, the iris setosa, versicolor and virginica. For this classification the last feature of the dataset was dropped to map to the three inputs architecture. Accuracies of upwards 96.6\% were attained for this task. 

\subsection{MNIST Handwritten digit classification}
The previous two case studies do not benefit significantly from the temporal dynamics of the implemented recurrent spiking neural network. Hence the final case study is the MNIST but in Sequential Order. This was tested with and without recurrency enabled. For this case study, as can be seen in Fig. \ref{fig:neural1}, the first three digits of the MNIST dataset (zero, one and two) were chosen. Additionally to take full advantage of the implemented architecture the images of the dataset are split into three equal segments, the top , middle and bottom part. In order for those parts to be split equally the images were distorted and additional padding pixels were added in the images.  The last step facilitated the reduction of the time-steps needed from 784 to 262. Those three segments were provided as inputs for the three input-neurons of the RSNN network and accuracies of 89\% were obtained for the multi-class classification (zero,one and two).  Binary classification yielded 94.7\%. The latter case study showcased the potential of the spiking and recurrent spiking neural networks.

\section{Results and Discussion}

ChatGPT has successfully generated the Verilog hardware description code for each module after a total of 117 iterations.  Table \ref{tab:tab_results} details the iterative conversational design process with ChatGPT for each hardware module. The LIF Neuron module and RLIF Layer proved to be the most challenging, requiring 38 and 17 iterations respectively, across 2 separate conversations.

Two factors contributed to the larger number of iterations: the complex overflow/underflow management, with bitwidth adjusting to adequately handle input current accumulations, and in the number of lines of the code. A module with fewer lines were easier to troubleshoot, as seen with the RLIF neuron module, which had to handle overflow/underflow, but has fewer lines of code.

The findings highlight the importance of a modular system design and the need for clear and well-defined requirements, supplemented by practical examples.


\begin{table}[t]
\centering
\caption{Verilog Hardware Description Code Generation Results}
\begin{center}
\begin{tabular}{|l|l|l|l|l|l}
\cline{1-5}
\textbf{\begin{tabular}[c]{@{}l@{}}Module\\ Name\end{tabular}} & \textbf{\begin{tabular}[c]{@{}l@{}}Chat\\ \#\end{tabular}} & \textbf{\begin{tabular}[c]{@{}l@{}}Itera-\\ tions\\ \#\end{tabular}} & \textbf{\begin{tabular}[c]{@{}l@{}}Lines\\ count\\ \end{tabular}} & \textbf{Improvements Required} &  \\ \cline{1-5}
\begin{tabular}[c]{@{}l@{}}LIF\\ Neuron\end{tabular} & 2 & \begin{tabular}[c]{@{}l@{}}38\\ (24+14)\end{tabular} & 33 & \begin{tabular}[c]{@{}l@{}}- Overflow/underflow\\    management\\    (Bit Width Adjustments +\\    sign extension)\\ - Verilog Syntax +\\    best practices\\ - Adding proper inputs\end{tabular} &  \\ \cline{1-5}
\begin{tabular}[c]{@{}l@{}}RLIF\\ Neuron\end{tabular} & 1 & 6 & 19 & \begin{tabular}[c]{@{}l@{}}- Overflow/underflow\\    management\\   (Bit Width Adjustments\\   + sign extension)\end{tabular} &  \\ \cline{1-5}
\begin{tabular}[c]{@{}l@{}}RLIF\\ Layer\end{tabular} & 2 & \begin{tabular}[c]{@{}l@{}}17\\ (10+7)\end{tabular} & 63 & \begin{tabular}[c]{@{}l@{}}- Clarification of requirements\\   (format of input data, \\   parameter sharing among\\   neurons, behaviour of the\\   module)\\ - Verilog Syntax\\ - Oveflow/underflow\\    management\end{tabular} &  \\ \cline{1-5}
RSNN & 1 & 8 & 37 & \begin{tabular}[c]{@{}l@{}}- Clarification on module\\    behaviour\\ - Verilog best practices\end{tabular} &  \\ \cline{1-5}
\begin{tabular}[c]{@{}l@{}}FIPO\\ Memory\end{tabular} & 1 & 12 & 21 & \begin{tabular}[c]{@{}l@{}}- Clarification on module\\   behaviour\\ - Adding control input signals\end{tabular} &  \\ \cline{1-5}
RegN & 1 & 5 & 6 & \begin{tabular}[c]{@{}l@{}}- Module Parameterizability\\ - Adding control input signals\\ - Change in reset behaviour\end{tabular} &  \\ \cline{1-5}
\begin{tabular}[c]{@{}l@{}}Control\\ Memory\end{tabular} & 1 & 9 & 55 & \begin{tabular}[c]{@{}l@{}}- Clarification on model\\   behaviour\\ - Verilog Syntax\end{tabular} &  \\ \cline{1-5}
\begin{tabular}[c]{@{}l@{}}Top\\ Module\end{tabular} & 1 & 22 & 103 & \begin{tabular}[c]{@{}l@{}}- Clarification on model\\   behaviour\\ - Refining Connections\end{tabular} &  \\ \cline{1-5}
\end{tabular}
\label{tab:tab_results}
\end{center}
\end{table}

\subsection{FPGA Prototyping Results}
For the FPGA prototyping, the Xilinx Vivado Design Suite along with the Digilent CMOD S7 board, which is equipped with the Xilinx Spartan 7 FPGA. The post-implementation results of the ChatGPT-generated design show an utilization of 1011 LUTs and 507 FFs.
The maximum allowed system clock frequency is 83 MHz. Power analysis was conducted using the Switching Activity Interchange Format (SAIF) file generated during a timing post-implementation simulation and revealed a total power consumption of 65 mW, of which 4 mW is dynamic power and 61 mW static power.

\begin{figure}[t]
\centering
\centerline{\includegraphics[width=\columnwidth ]{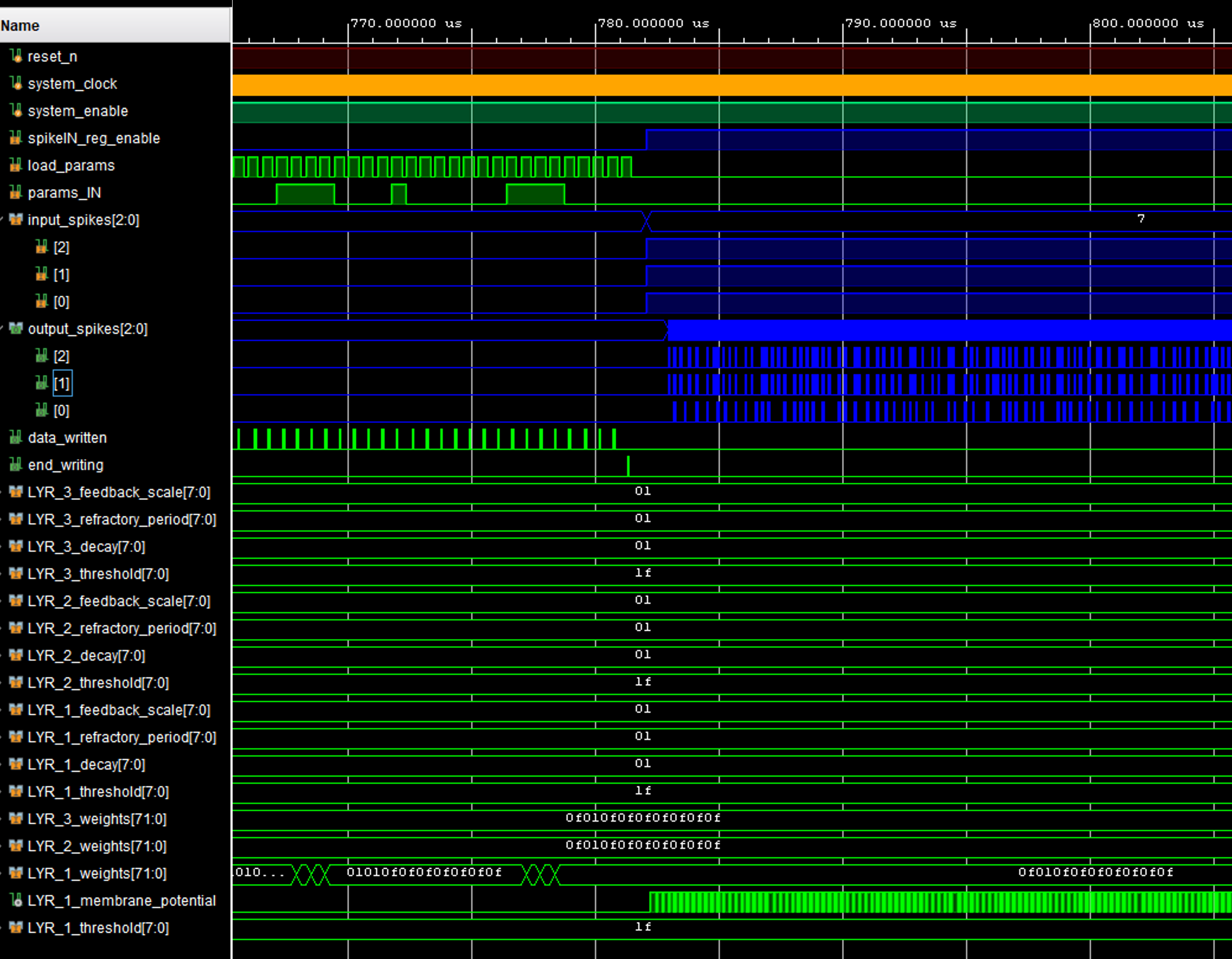}}
\caption{Timing Post-Implementation Simulation of the RSNN hardware design generated by ChatGPT on Spartan 7 FPGA. The system operates in two functional modes: startup mode, where load\_params goes high enabling the loading of all network parameters into memory, and running mode, where the spikeIN\_reg\_enable signal enables the storage of input spikes and their processing, allowing the system to generate output spikes.}
\label{fig:simulation_1}
\end{figure}

\begin{figure}[t]
\centering
\centerline{\includegraphics[width=\columnwidth ]{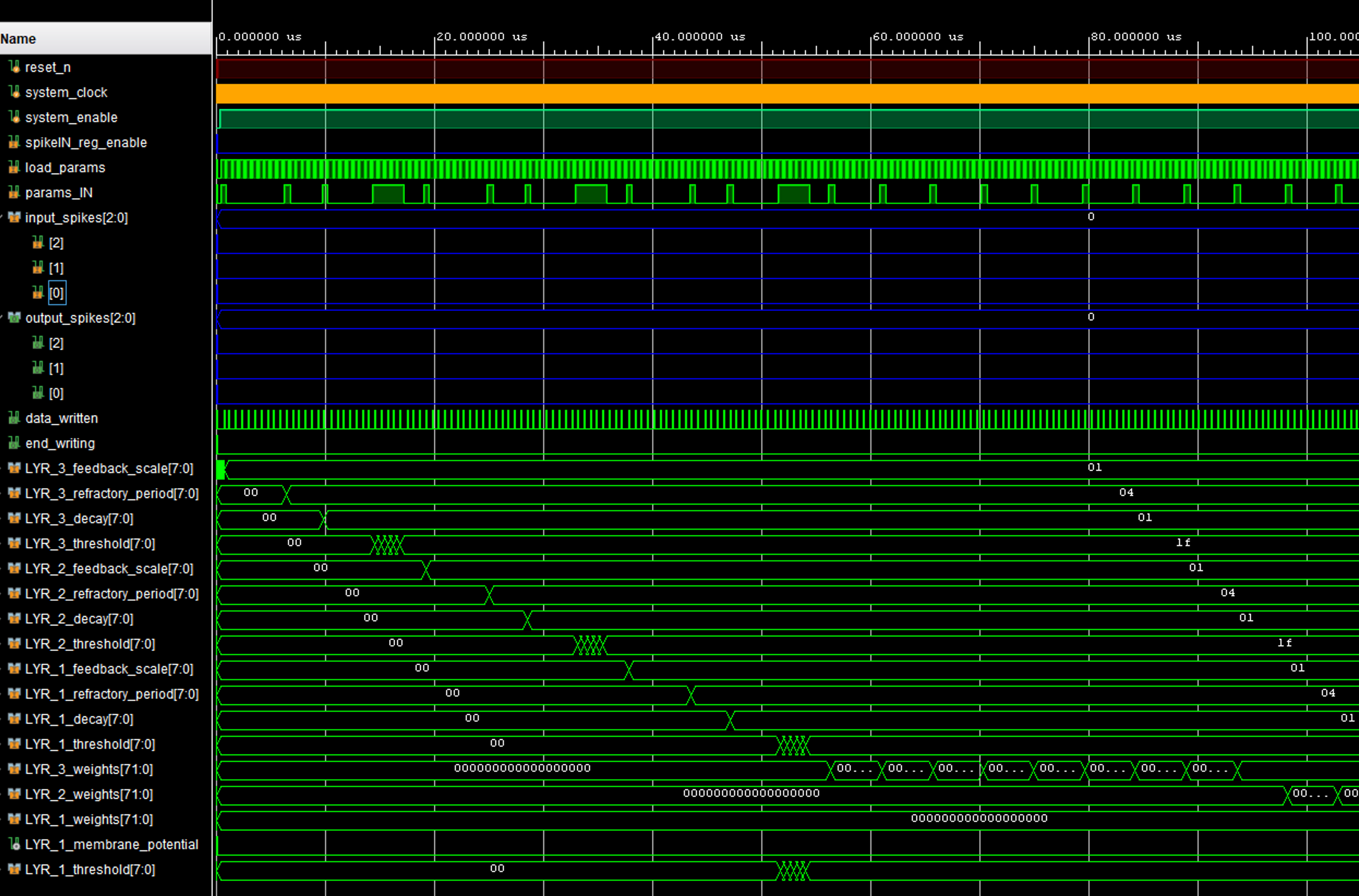}}
\caption{Timing Post-Implementation Simulation of the RSNN hardware design generated by ChatGPT on Spartan 7 FPGA. Zoomed-in view of the startup mode, where the load\_params signal is asserted high, enabling the loading of all network parameters into memory. This phase is crucial for initializing the network with the correct parameters before transitioning to the running mode.}
\label{fig:simulation_2_startup}
\end{figure}

Fig. \ref{fig:simulation_1} and \ref{fig:simulation_2_startup} show a timing post-implementation simulation. It can be seen that the system has two functional modes: startup mode and running mode.
In the initial mode, all the network parameters are loaded into memory while in the second the input spikes are processed. 

\subsection{ASIC implementation and Tiny Tapeout}
The  design was implemented using SkyWater 130 nm technology and OpenLane, an open-source EDA flow. OpenLane automates the digital hardware design steps, from RTL synthesis and implementation to the generation of the Graphic Design System II (GDSII) file, which is the final output containing all the layout information.

\begin{figure}[t]
\centerline{\includegraphics[width=\columnwidth]{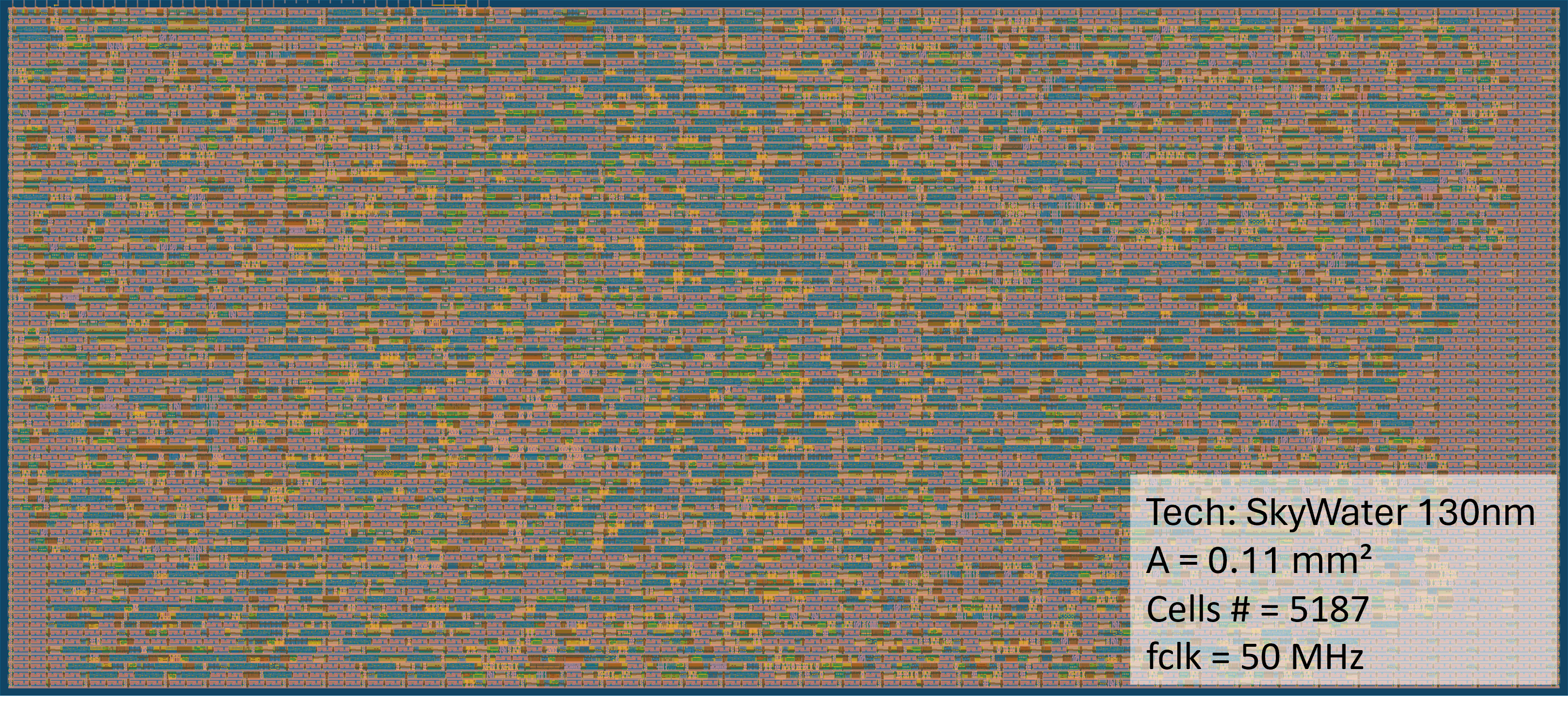}}
\caption{2D Visualization of the GDSII file of the RSNN hardware design generated by ChatGPT in SkyWater 130 nm technology.}
\label{fig:gds}
\end{figure}

Fig. \ref{fig:gds} presents a 2D visualization of the final GDSII file. The implementation results show an area occupation of 0.11 mm$^2$ and 5187 total cells, distributed as follows: 1635 combo logics, 1577 taps, 647 NORs, 580 ORs, 512 FFs, 495 buffers, 345 NANDs, 337 ANDs, 242 Misc, 220 multiplexers, 124 inverters, 49 diodes).

These results, combined with the all signoff tests passed, demonstrate the robustness  of the hardware design asissted by ChatGPT and highlight its potential as a useful tool in the hardware design process.

\section{Conclusions}
In this paper, the hardware architecture of a programmable RSNN generated by the ChatGPT LLM was presented. The design was verified through FPGA prototype and also synthesized and taped out in the SkyWater 130 nm technology -TinyTapeout 6-.

The system was evaluated by performing inference  with three applications, the exclusive OR, the IRIS flower classification, and hand-written digit classification MNIST, obtaining an accuracy upwards 96\%. 

Future work will aim to explore the generation of hardware description code for more complex networks and to build more comprehensive test benches as well as training ``Small Language Models'' using Verilog training data.

\section*{Acknowledgment}
This work was supported by NSF Grant 2020624 AccelNet:Accelerating Research on Neuromorphic Perception, Action, and Cognition through the Telluride Workshop on Neuromorphic Cognition Engineering, NSF Grant 2332166 RCN-SC: Research Coordination Network for Neuromorphic Integrated Circuits and NSF Grant 2223725 EFRI BRAID: Using Proto-Object Based Saliency Inspired By Cortical Local Circuits to Limit the Hypothesis Space for Deep Learning Models.

\bibliographystyle{IEEEtran}
\bibliography{IEEEabrv,bibliography}

\end{document}